\documentclass[aps,prb,twocolumn,showpacs,superscriptaddress,groupedaddress]{revtex4}  
\usepackage{graphicx}
\usepackage{amsmath}
\usepackage{dcolumn}
\usepackage{bm}

\begin{document}


\title{Finite momentum superconductivity in superconducting hybrids: Orbital mechanism}

\author{M. Yu. Levichev}
\affiliation{Institute for Physics of Microstructures, Russian
Academy of Sciences, 603950, Nizhny Novgorod, GSP-105, Russia}

\author{I. Yu. Pashenkin}
\affiliation{Institute for Physics of Microstructures, Russian
Academy of Sciences, 603950, Nizhny Novgorod, GSP-105, Russia}

\author{N. S. Gusev}
\affiliation{Institute for Physics of Microstructures, Russian
Academy of Sciences, 603950, Nizhny Novgorod, GSP-105, Russia}

\author{D.Yu. Vodolazov}

\email{vodolazov@ipmras.ru}

\affiliation{Institute for Physics of Microstructures, Russian
Academy of Sciences, 603950, Nizhny Novgorod, GSP-105, Russia}

\date{\today}

\begin{abstract}

Normally in superconductors, as in conductors, in the state with
zero current $I$ the momentum of superconducting electrons $\hbar
q =0$. Here we demonstrate theoretically and present experimental
evidences that in superconducting/normal metal (SN) hybrid strip
placed in in-plane magnetic field $B_{in}$ finite momentum state
($\hbar q \neq 0$) is realized when $I=0$. This state is
characterized by current-momentum dependence $I(q)\neq -I(-q)$,
nonreciprocal kinetic inductance $L_k(I) \neq L_k(-I)$ and
different values of depairing currents $I_{dep}^{\pm}$ flowing
along the SN strip in opposite directions. Found properties have
{\it orbital} nature and are originated from gradient of density
of superconducting electrons $\nabla n$ across the thickness of SN
strip and field induced Meissner currents. We argue that this type
of finite momentum state should be rather general phenomena in
superconducting structures with artificial or intrinsic
inhomogeneities.
\end{abstract}

\maketitle

\section{Introduction}

Normally in superconductors, also as in normal conductors (metals
or semiconductors) the state with zero total current $I=0$ is
characterized by zero momentum $\hbar q=0$ of superconducting
electrons ($q=\nabla \phi+2\pi A/\Phi_0$, where $\phi$ is a phase
of superconducting order parameter, A is a vector potential and
$\Phi_0=\pi \hbar c/|e|$ is magnetic flux quantum). Such an {\it
ordinary} superconductor has antisymmetric current-momentum
dependence $I(q)=-I(-q)$ and symmetric kinetic inductance
$L_k(I)=L_k(-I)$ (see Fig. 1(a,d)). Kinetic inductance $L_k\sim
-dq/dI$ is a measure of inertia of superconducting electrons
possessing the kinetic energy $E_k \sim \int n\hbar^2q^2/(2m)dV
\sim \int L_k(I)IdI$ ($m$ is a mass of superconducting electrons,
$n$ is their density and V is a volume of superconductor). In
superconductors $L_k$ contributes to total inductance $L=L_k+L_g$,
where $L_g$ is ordinary (geometric) inductance, which does not
depend on $I$.
\begin{figure}[hbtp]
\includegraphics[width=0.5\textwidth]{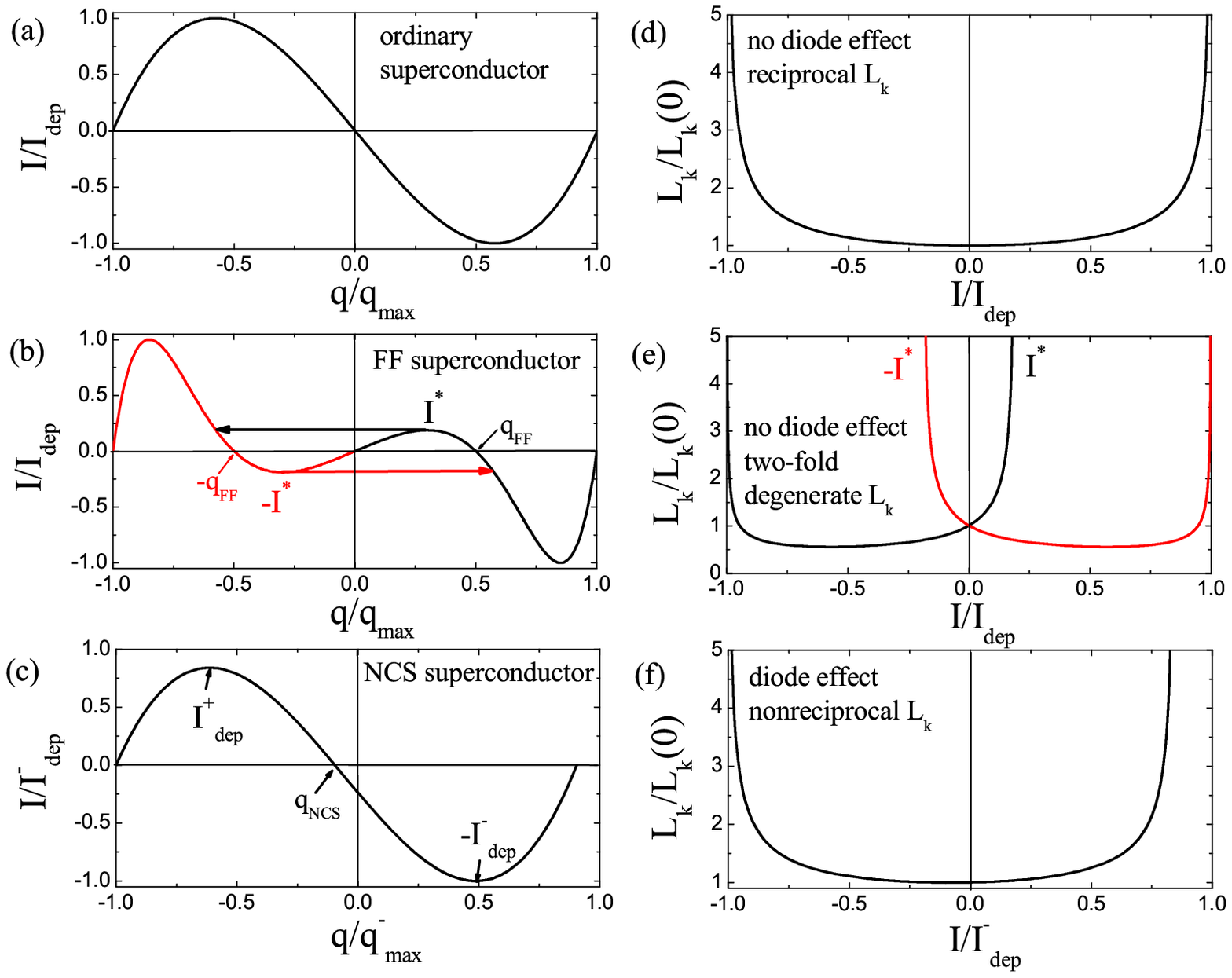}
\caption{Current-momentum dependence in ordinary (panel a),
Fulde-Ferrell (b) and noncentrosymmetric (c) thin and narrow
superconducting strip with uniform current distribution over the
width of superconductor. In FF strip $I(q)=-I(-q)$ (as in ordinary
superconductor) and there are two degenerate finite momentum
states ($\pm q_{FF}$) while in NCS strip $I(q)\neq -I(-q)$ and
there is one finite momentum state with $q=q_{NCS}$. In panels
(d-e) we present corresponding dependencies of kinetic inductance
$L_k(I)\sim -dq/dI$. In NCS superconductor $L_k(I)\neq L_k(-I)$
which is a fingerprint of this state, also as the diode effect
($I_{dep}^-\neq I_{dep}^+$). In ordinary and FF superconductors
$I_{dep}^-=I_{dep}^+$, while $L_k(I)$ is two fold degenerate in FF
superconductor in current range $-I^*<I<I^*$.}
\end{figure}

However it has been predicted $q\neq 0$ despite of $I=0$ for some
superconducting systems. The most familiar examples are the
ferromagnetic superconductor with magnetic exchange energy of
order of superconducting gap $\Delta$ and thin superconducting
strip placed in large in-plane magnetic field $\mu_B B_{in} \sim
\Delta$. In these systems superconducting pairing occurs with
finite center-of-mass momentum of electrons $q_{FF}$ as it has
been shown by Fulde and Ferrell \cite{Fulde_1964} due to Zeeman
splitting of energy of electrons having opposite spin. Recently,
the similar Fulde-Ferrell (FF) state has been found theoretically
in superconductor/ferromagnet (SF) \cite{Mironov_2012,JETPL-2014},
superconductor/ferromagnet/normal metal (SFN) \cite{Mironov_2018}
and nonequilibrium SN hybrids
\cite{Bobkova_2013,Ouassou_2020,Levichev_2021}. In these hybrid
superconductors there are local nonzero currents flowing in
opposite directions in different layers which distinguishes it
from the original FF state. Both cases we call here as FF
state/superconductor. FF superconducting strip at $I=0$ has two
degenerate states with $q=\pm q_{FF}$ due to finite size effect
\cite{Plastovets_2019} (for infinite sample $q_{FF}$ may have any
direction) and antisymmetric $I(q)=-I(-q)$ dependence
\cite{Samokhin_2017,Plastovets_2019} (see Fig. 1(b)). At small
currents $-I^*<I<I^*$ this system has two stable states
\cite{Samokhin_2017} which have different values of kinetic
inductance $L_k$ \cite{Marychev_2021} - see Fig. 1(e). Depairing
current in FF superconductor does not depend on current direction
which reflects absence of particular direction for $q_{FF}$. As
current increases and approaches to $\pm I^*$ (see Fig. 1(b)) FF
superconductor switches to the state having opposite $q_{FF}$ at
$I=0$. If during this transition FF superconductor is not heated
considerably it stays in superconducting state
\cite{Plastovets_2019}.

Besides that there is another class of superconducting materials
where finite momentum superconductivity may exist. These are so
called noncentrosymmetric (NCS) superconductors with no inversion
center and where in presence of spin-orbit coupling and in-plane
magnetic field finite momentum of superconducting electrons
$q_{NCS}$ with {\it particular} direction appears
\cite{Edelshtein_JETP_1989,Agterberg_2003,book_2012}. It has been
discussed recently that such a NCS superconductor should have
different depairing (critical) currents flowing either in parallel
or antiparallel to $\vec{q}_{NCS}$
\cite{Yuan_PNAS_2022,Daido_PRL_2022,He_NJP_2022} and nonreciprocal
kinetic inductance $L_k(I)\neq L_k(-I)$
\cite{Baumgartner_Nat_Nan_2022} following from the
current-momentum dependence $I(q)\neq -I(-q)$ (see Fig. 1(c,f)).
Difference between critical currents means that ac current with an
amplitude between these critical values should produce in NCS
superconductor voltage of only one sign. This is the reason why
the difference between critical currents is called as a
superconducting diode effect (SDE).

In our work we present theoretical and experimental results which
demonstrate that finite momentum superconductivity (FMS), typical
for discussed above noncentrosymmetric superconductors, is
realized in the supercondutor/normal metal hybrid {\it without}
spin-orbit coupling in presence of in-plane magnetic field (see
Fig. 2(a)). In SN bilayer there is a thickness dependent 'density'
of superconducting electrons $n(z)$ which is a coefficient between
superconducting current density and momentum:
$j(z)\sim-|e|n(z)q(z)$. In N layer finite $n$ appears due to
proximity induced superconductivity, and usually it is smaller
than in S layer (this case is shown in Fig. 2(b)). In-plane
magnetic field induces Meissner currents $j(z)$ and
superconducting electrons posses momentum $q(z)$. For thin SN
bilayer with thickness $d_S+d_N \ll \lambda$ ($\lambda \sim
n^{-1/2}$ is a London penetration depth) one may neglect magnetic
field arising from the Meissner currents and choose ${\bf
A}=(B_{in}z,0,0)$ where $-(d_S+d_N)/2<z<(d_S+d_N)/2$. In case
$n(z) = const$ the total current $I\sim \int j(z)dz=0$ when the
{\it thickness averaged momentum} $q_0=\int q dz/(d_S+d_N)=\nabla
\phi=0$. But in case $\nabla n(z) \neq 0$ one needs finite
$q_0=q_{NCS}$ to have $I=0$. It describes the {\it orbital}
mechanism of finite momentum superconductivity in SN strip. From
mathematical point of view presence of both thickness averaged
$\nabla n$ and $B_{in}$ breaks the symmetry in SN strip and vector
$\nabla n \times B_{in}$ defines the particular direction and
value of $q_{NCS}$.

\begin{figure}[hbtp]
\includegraphics[width=0.48\textwidth]{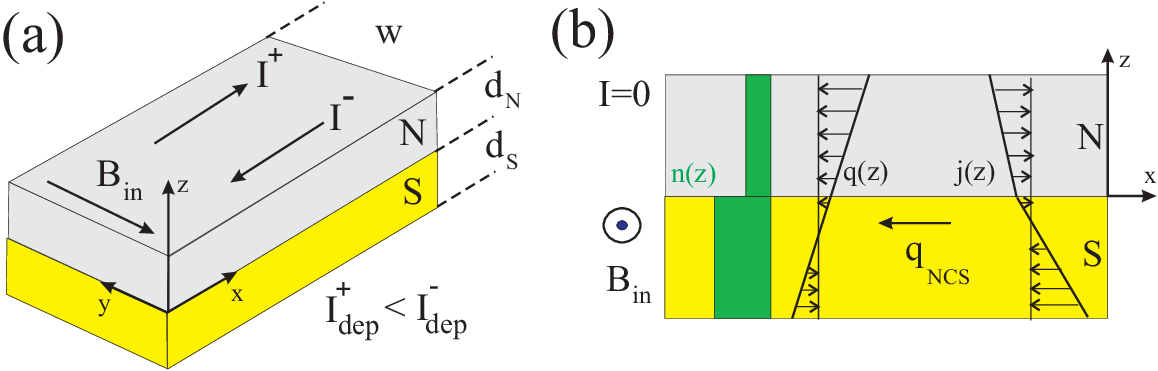}
\caption{(a) SN strip with transport currents of different
directions placed in the in-plane magnetic field. (b) Sketch of
thickness dependent 'density' of superconducting electrons $n(z)$,
momentum $q(z)$ and superconducting current density
$j(z)\sim-|e|n(z)q(z)$ in SN strip. For shown $n(z)$ and direction
of $B_{in}$ finite momentum $q_{NCS}$ points against axis x.}
\end{figure}

From above consideration it is clear that finite momentum state of
this type has to exist in any superconductor having finite $\nabla
n$ and $B_{in}$. Here we prove it for SN bilayer where S layer is
a dirty superconductor with large resistivity (small diffusion
coefficient $D_S$) in the normal state and N layer is a
low-resistive normal metal having large diffusion coefficient
$D_N\gg D_S$. In this system due to noticeable contribution of
proximity induced superconductivity in N layer to total $L_k$ we
expect to have large difference between $L_k(I)$ and $L_k(-I)$
which is easy to observe experimentally and which is a fingerprint
of FMS. This system also has diode effect, as it has been found
earlier in Ref. \cite{Vodolazov_2018} but its relation with FMS
was not discussed there. Another motivation to study this system
comes from recent experiments where diode effect was observed in
somewhere similar superconducting hybrids in presence of in-plane
magnetic field
\cite{Baumgartner_Nat_Nan_2022,Ando_Nat_2020,Turini_N_Lett_2022,Pal_Nat_Phys_2022,Sundaresh_2023}
and where we also expect contribution of the orbital mechanism to
SDE.

The structure of our paper is following. In section II we present
our theoretical results with characteristics of FMS which has an
orbital mechanism in SN bilayer. In section IIIA we describe used
experimental methods to study FMS in SN bilayer. In sections
IIIB,C we present experimental results which demonstrate that we
have finite momentum state in our superconducting hybrid and some
unexpected its properties. In section IV we discuss that orbital
mechanism of FMS has distinctive thickness dependence (it can be
used to distinguish it from the mechanism connected with
spin-orbit coupling) and our SN hybrid has expected behavior. In
the same section we discuss other experiments on FMS and diode
effect and its relation with our results. In section V we make
conclusions.

\section{Theoretical results}

We start with presentation of our theoretical results. In Fig. 3
we show calculated $I(q_0,B_{in})$ and $L_k(I,B_{in})$ for the SN
strip having following parameters: $d_S=d_N=4\xi_c$ ($\xi_c=(\hbar
D_S/k_BT_{c0})^{1/2}$, $T_{c0}$ is a critical temperature of S
layer) and ratio $D_N/D_S=100$. To find it we use Usadel model
(details of calculations are present in Appendix A). We choose two
temperatures $T=0.2$ and $T=0.6 T_{c0}$ which correspond to
different physical situations. At $T=0.2 T_{c0}$ contribution of N
layer is dominant in $L_k$ at small fields and currents while at
$T=0.6 T_{c0}$ N layer contributes much smaller in transport
properties. In both cases finite field-controlled $q_{NCS}$
appears (see upper insets in panels (a,b) of Fig. 3) leading to
asymmetry of $I(q_0)$ and nonreciprocal $L_k(I)$. At $T=0.2
T_{c0}$ and small $B_{in}$ $q_{NCS}>0$ because $n$ is larger in N
layer (in Fig. 2 opposite situation is shown, which is true for
our SN strip at $T=0.6T_{c0}$ at any $I$ and $B_{in}$). At large
$B_{in}$ and currents close to $I_{dep}^{\pm}$ superconductivity
in N layer becomes suppressed and $L_k$ increases. Transition
between the states with different $L_k$ is accompanied by
appearance of the peak in dependencies $L_k(I)$ and $L_k(B_{in})$
- see Fig. 3(c,d). Note that this peak is absent in ordinary
superconductors (see Fig. 1(d)) and/or much less pronounced in SN
bilayers with small contribution of N layer to transport
properties (see panel (d) in Fig. 3).

In the finite momentum state there is a difference between
positive and negative depairing currents (see bottom insets in
panels (a,b) of Fig. 3). $I_{dep}^-$ is larger than $I_{dep}^+$
because current induced momentum partially compensate field
induced momentum in N layer leading to recovery of proximity
induced superconductivity. At small $B_{in}$ it provides even
increase of $I_{dep}^-$ - the physically similar increase of $I_c$
is realized in superconducting strip with nonequivalent edges
being in out-of-plane magnetic field \cite{Vodolazov_2005}.

\begin{figure}[hbtp]
\includegraphics[width=0.48\textwidth]{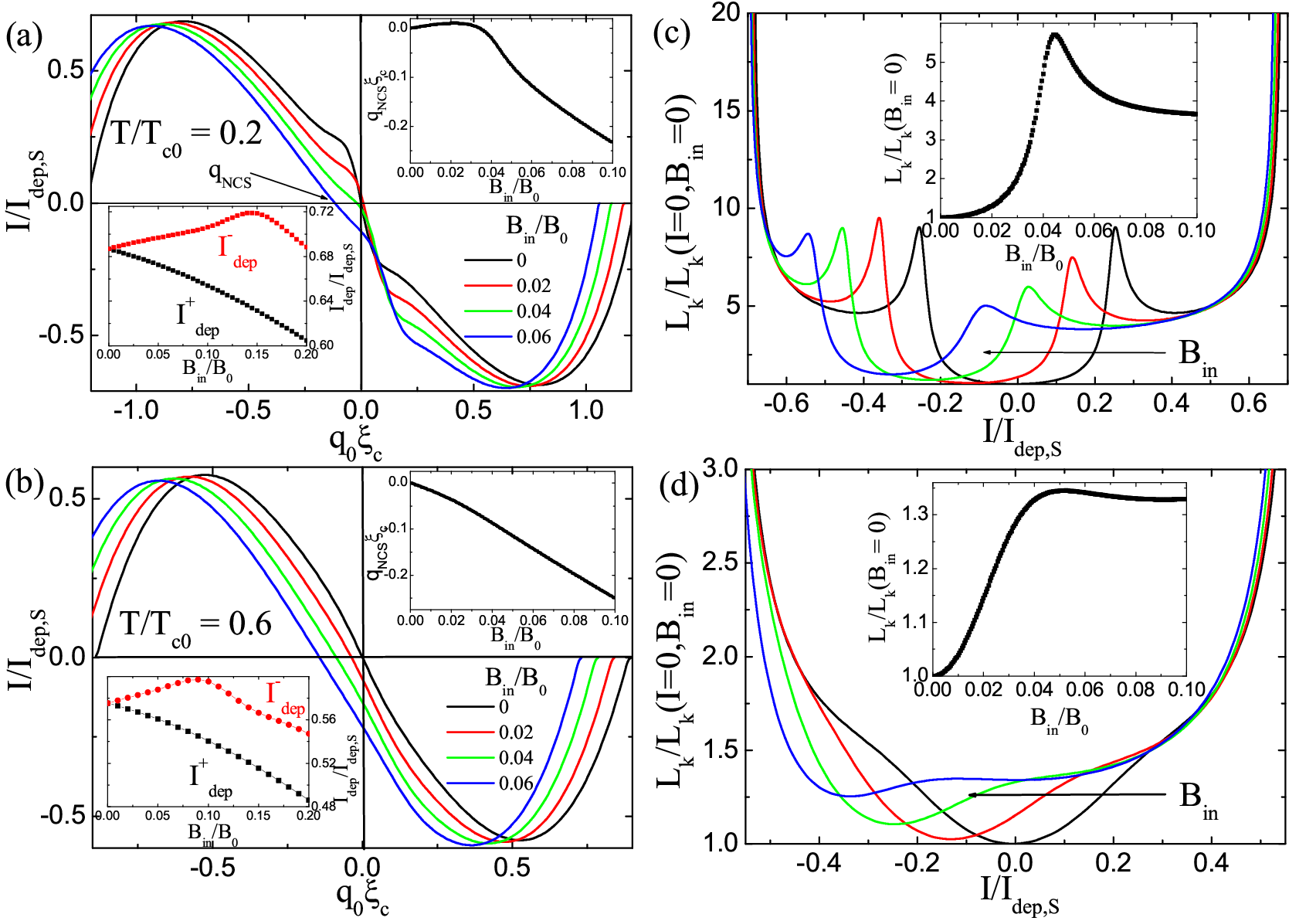}
\caption{Theoretical dependencies of current on momentum (a,b) and
kinetic inductance on current (c,d) in SN bilayer at different
in-plane magnetic fields and two temperatures $T=0.2$ and
$0.6T_{c0}$. In insets of panels (a,b) we present field
dependencies of $I_{dep}^{\pm}$ and finite momentum $q_{NCS}$ at
$I=0$. In insets of panels (c,d) we show calculated field
dependence of $L_k$ in zero current state. Here $B_0=\Phi_0/2\pi
\xi_c^2$ and $I_{dep,S}$ is a depairing current of single S layer
with thickness $d_S$.}
\end{figure}

\section{Experiment}

\subsection{Methods}

Experiment has been made with MoN/Cu strips. MoN is a dirty
superconductor with resistivity in the normal state $\rho=150\mu
Ohm \cdot cm$ while Cu is low resistive metal (40 nm thick Cu has
$\rho=2.4 \mu Ohm \cdot cm$ at T=10 K). The MoN/Cu bilayers are
grown by magnetron sputtering with a base vacuum level of the
order of $1.5 \cdot 10^{-7} {\rm mbar}$ on standard silicon
substrates without removing the oxide layer and at room
temperature. At first Mo is deposed in an atmosphere of a gas
mixture Ar : N$_2$ = $10 : 1$ at a pressure of $ 1 \cdot 10^{-3}
\rm{mbar}$ and than Cu is deposed in an argon atmosphere at a
pressure of $1\cdot 10^{-3} {\rm mbar}$. Finally, MoN/Cu strips
has been made with help of mask free optical lithography.

Majority of measurements are done for thick $d_{MoN}=40 nm$,
$d_{Cu}=40 nm$ samples. Altogether we have four long strips A1-A4
(width $4 \mu m$, length $3 mm$) and several short bridges B1-B9
(width $4 \mu m$, length 100 $\mu m$). All measured samples
(A1-A2,A4, B2-B3) have nearly the same sheet resistance
$R_s(300K)=1 \Omega$ and $R_s(T=10K)=0.6 \Omega$ (variation from
sample to sample is less than $10\%$) and critical temperature
$T_c=7.87K$ defined from condition that resistance $R(T_c)=0.5
R(10K)$.

In addition we have made and studied two times thinner
(MoN(20nm)/Cu(20nm)) reference long strips with the same width and
length as thicker MoN/Cu strips to check the thickness dependence
of the finite momentum state and verify its orbital nature. Their
characteristics are present in Appendix C.

\begin{figure}[hbtp]
\includegraphics[width=0.48\textwidth]{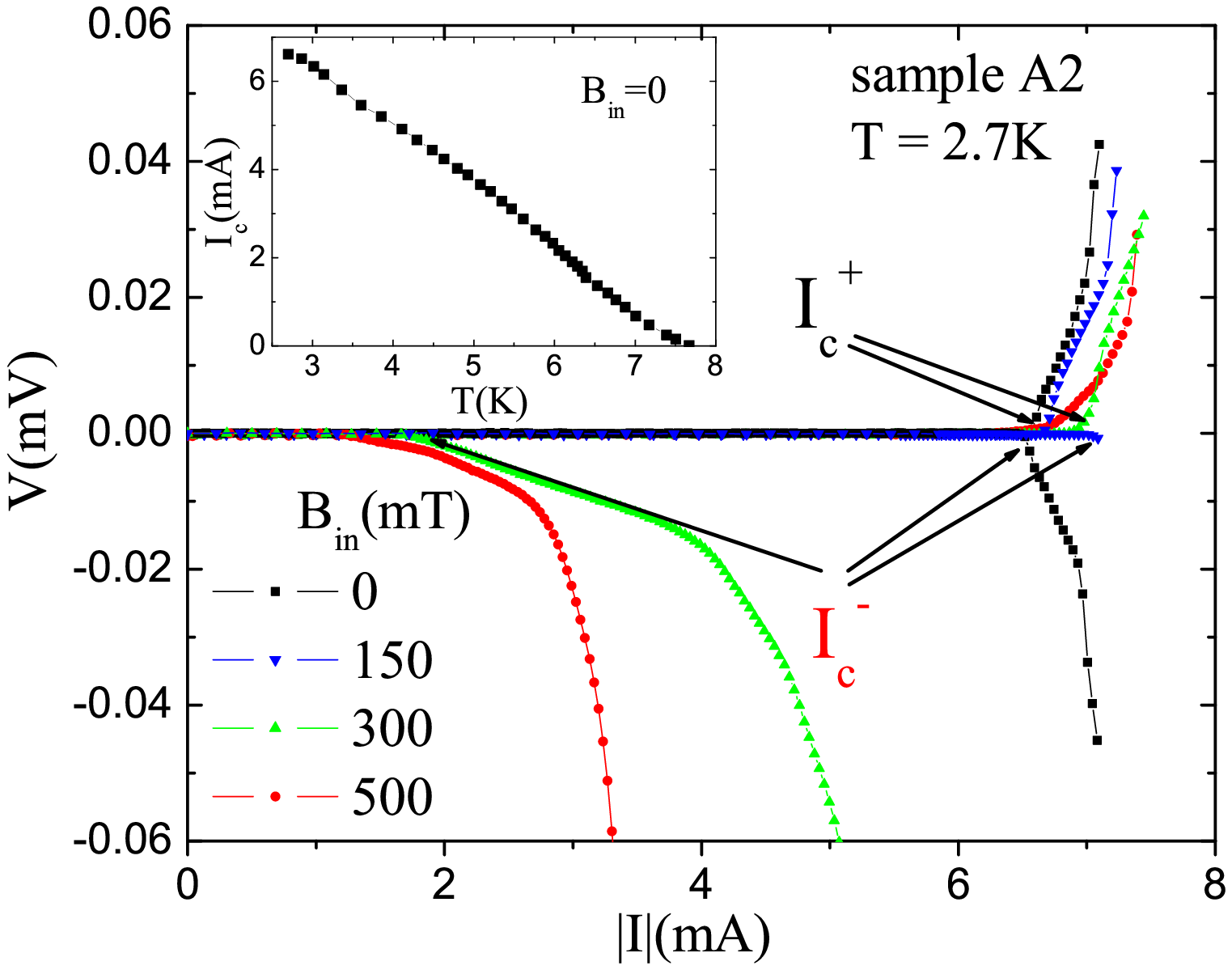}
\caption{Current-voltage characteristics of sample A2 at different
$B_{in}$ and $T=2.7K$ (negative voltage corresponds to negative
current while in x axis we show absolute value of the current),
arrows indicate $I_c^{\pm}$. In the last points of IV curves there
is jump to the normal state (not shown here). IV curves are not
hysteretic before transition to the normal state. Critical
currents (their absolute values) are determined from condition
$|V|(I=I_c^{\pm}) = 0.5 \mu V$. After transition to the normal
state sample returns to the superconducting state at $I_r \simeq
1.6 mA$ (when $I_c^{\pm}>I_r$) which practically does not depend
on magnetic field and current direction. In inset we show
temperature dependence $I_c(T)$ at $B_{in}=0$.}
\end{figure}

Current voltage characteristics has been measured by four-probe
method. Examples of IV curves are shown in Fig. 4 for sample A2
(in inset we present temperature dependence of $I_c$ at
$B_{in}=0$). From these measurements we extracted critical
currents $I_c^{\pm}$ (see arrows in Fig. 4) as function of
$B_{in}$.

Impedance $Z=Z_{re}+iZ_{im}$ of MoN/Cu strip has been measured
using Standford Research SR830 lock-in amplifier. For measurements
the four-probe method was used. Excitation signal with frequency
$\nu = 100kHz$ and voltage 50mV from SR830 internal generator was
supplied through 1 $k\Omega$ resistor. Signal from the sample is
applied to the differential input of the SR830 through central
electrodes of two coaxial cables. These measurements allow us to
find field and current dependence of inductance $L=Z_{im}/{2\pi
\nu}$. The same device has been used to measure first
($R_{\omega}$) and second ($R_{2\omega}$) harmonic signals of the
ac resistance at frequency of 130 Hz.

In our strips we have noticeable contribution of geometric
inductance $L_g$ to the total inductance. Using expression below
for the strip with length $l$, width $w$ and thickness $d$ in
limit $l \gg (w+d)$ \cite{Greenhouse}

\begin{equation}
L_g=\frac{\mu_0l}{2\pi}\left(\ln\left(\frac{2l}{w+d}\right)+1/2\right)
\end{equation}

we find for our long strips $L_g \simeq 4.7 nH$.

\subsection{Results below T$_c$}

In Fig. 5 (see also Fig. 8 in Appendix B) we present $L_k(I)$ at
different $B_{in}$ and $I_c{\pm}(B_{in})$ at two temperatures 4.6
and 2.7 K which roughly corresponds to two temperatures present in
Fig. 3. The main result is that in in-plane field inductance $L$
is nonreciprocal and it is the {\it main} proof that MoN/Cu strip
has finite momentum at $I=0$. Although we know $L(I,B_{in})$ it
does not allow us to find $q_{NCS}$ and its dependence on magnetic
field - for that one has to know additionally $I(q_0)$ at least at
one value of $q_0$.

\begin{figure}[hbtp]
\includegraphics[width=0.48\textwidth]{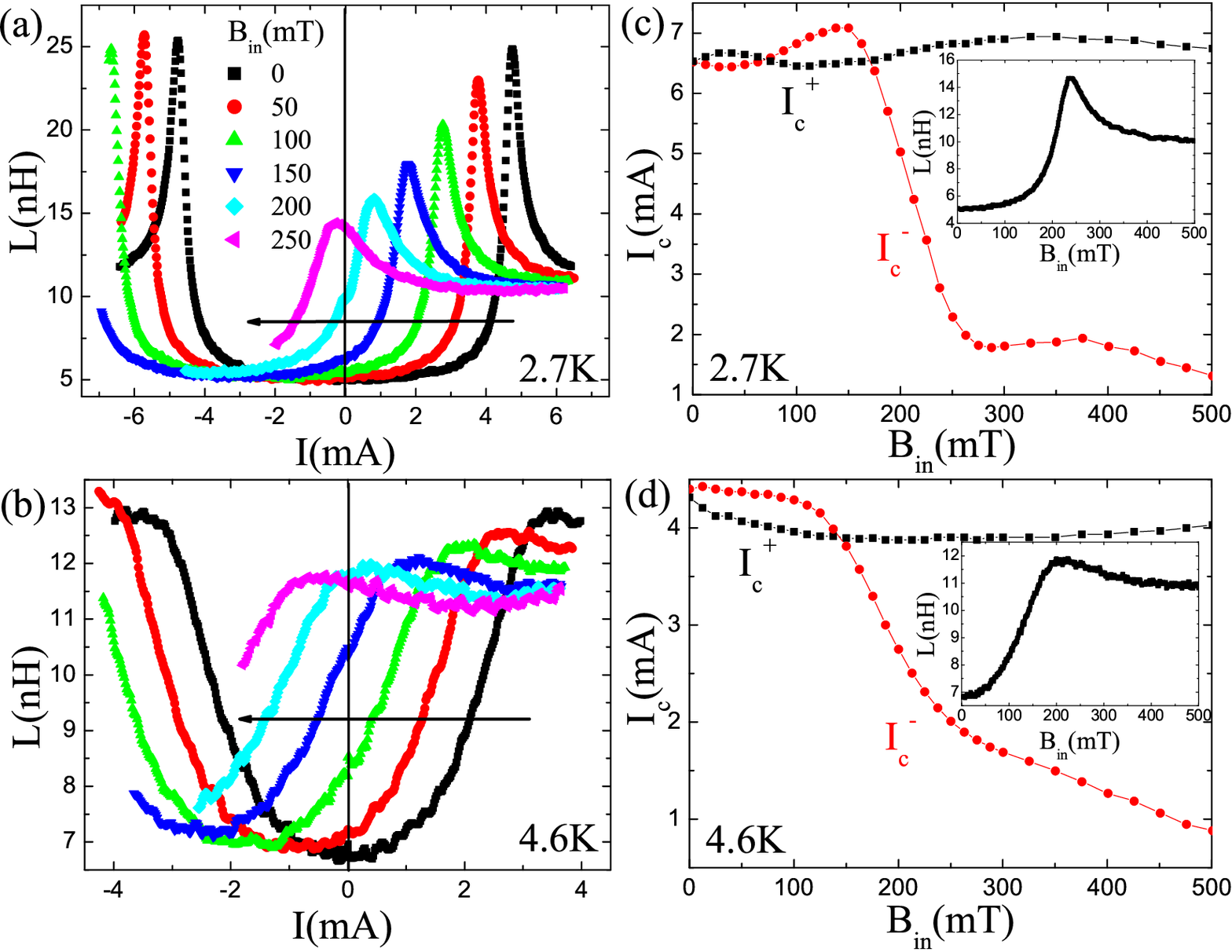}
\caption{Panels (a,b) - evolution of current dependent inductance
of MoN/Cu strip (sample A2) in superconducting state with increase
of in-plane magnetic field at T=2.7K and T=4.6K, respectively. In
panels (c,d) we present field dependent critical current and field
dependent inductance at $I=0$.}
\end{figure}

Experimental dependence $L(B_{in})$ at zero current and evolution
of $L(I)$ with increase of $B_{in}$ follow theoretical prediction
at all magnetic fields with the important difference, for
following discussion of diode effect, that in the experiment we do
not reach the depairing current $I_{dep}^{\pm}$ where theoretical
$L_k$ diverges (compare Figs. 5(a,b) with Figs. 3(c,d)). It occurs
probably due to presence of edge defects which allow entry of {\it
out-of-plane} vorticies at $I<I_{dep}^{\pm}$ and it does not allow
us to approach the depairing current. Indeed, image of the similar
Cu/MoN strip made with help of electron microscope (see for
example Fig. 1 in \cite{JETP_Ustavschikov}) shows that we have
edge roughnesses with size of about 100 nm. Idea about vortex
entry at $I>I_c^{\pm}$ is supported by experimental IV curves (see
Fig. 4) - above the critical current SN strip transits to low
resistive state which resembles flux flow regime.

In contrast to inductance our experimental results on
superconducting diode effect are controversial. We find the
sample-dependent difference between $I_c^+$ and $I_c^-$ (in
contrast to almost not sample-dependent $L(I,B_{in})$) and besides
it has unexpected value and sign at relatively large $B_{in}$ for
all studied samples (compare Figs. 3, 5, 8 and 9). At magnetic
field $B_{in} \lesssim 150 mT$ the sign and value of SDE mainly
coincide with prediction of our theory for samples A1, A4, B4
while for samples A2, B3 there is a sign change of diode effect
with increase of $B_{in}$. At $B_{in} \gtrsim 150 mT$ we have
$I_c^- \ll I_c^+$  for all samples. The large difference (ratio
$\eta=2|I_c^--I_c^+|/(I_c^-+I_c^+)>1$) and its 'wrong' sign cannot
be explained by our theory which predicts $I_c^-=I_{dep}^- \gtrsim
I_c^+=I_{dep}^+$.

\begin{figure}[hbtp]
\includegraphics[width=0.48\textwidth]{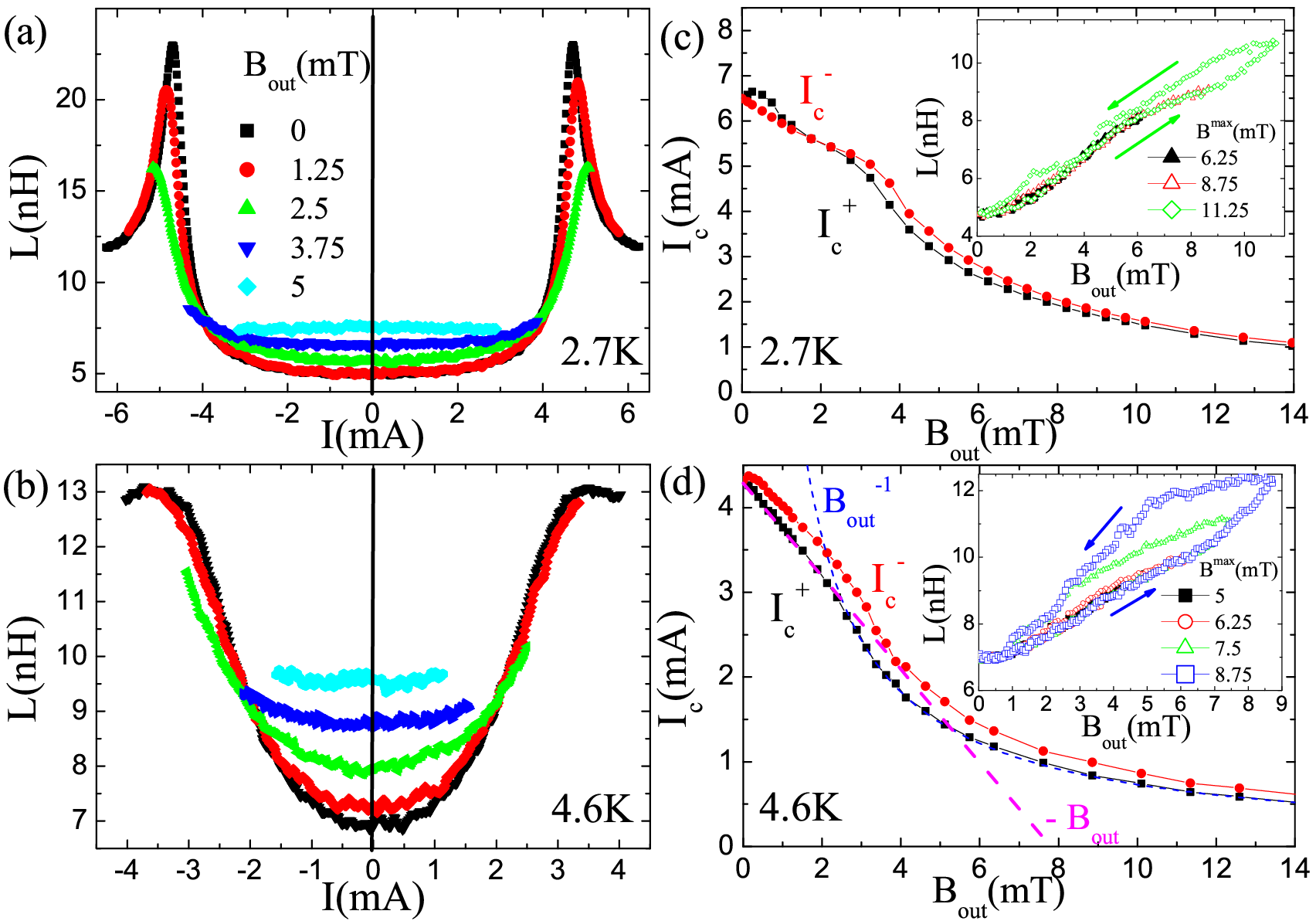}
\caption{Panels (a,b) - evolution of current dependent inductance
of MoN/Cu strip (sample A2) with increase of out-of-plane magnetic
field at T=2.7K and T=4.6K, respectively. In panels (c,d) we
present field dependent critical currents and inductance.
Dependence $L(B_{out})$ for sweeping up and down field is
hysteretic for relatively large amplitude of $B_{out}$ which we
relate with trapped out-of-plane vortices due to edge barrier for
vortex entry/exit.}
\end{figure}

It could be supposed that in our experiment together with in-plane
field there is small out-of-plane field $B_{out}$ (due to not
perfect alignment of the sample holder along superconducting coil
which is a source of our magnetic field). Even small $B_{out}$, in
comparison with $B_{in}$, strongly suppresses $I_c$ and in
presence of edge defects may lead to SDE which also has an orbital
nature because it originates from the combined effect of Meissner
and transport currents \cite{Vodolazov_2005}. Experimentally it
was found in different superconductors
\cite{Cerbu_NJP_2013,PRB14Ilin,Suri_APL_2022,Hou_2022,JETP_Ustavschikov,Satchell}.
Note, that probably the same mechanism is responsible for SDE
observed in NbSe$_2$ bridge \cite{Bauriedl_Nat_Comm_2022} which
follows from its sample-dependent character and nearly linear
decay of $I_c$ at small fields $\lesssim \Phi_0/2\pi \xi w$ which
points out on the edge barrier controlled $I_c^{\pm}$
\cite{Plourde_PRB_2001}. In the strip/bridge with edge defect
strength of this type of SDE is controlled by parameters of the
defect and $\eta$ may be larger than unity as it follows from Ref.
\cite{JETP_Ustavschikov} which resembles our result at large
$B_{in}$.

Therefore we measured $L(I,B_{out})$ and $I_c^{\pm}(B_{out})$ -
see Fig. 6. Influence of bulk pinning in our hybrid is negligible,
at least in used field range, as it follows from the typical for
edge-barrier controlled field-dependent $I_c^{\pm}(B_{out})$
\cite{Plourde_PRB_2001} ($I_c^{\pm}$ drops nearly linearly at low
fields and $I_c^{\pm}\sim 1/B_{out}$ at large fields - see panel
(d) in Fig. 6). We find nearly reciprocal $L(I)$ while there is
small superconducting diode effect, comparable in value with SDE
at $B_{in} \lesssim 150 mT$ and which also may change the sign
(see Fig. 6(c)). The sign change does not follow from the model of
Ref. \cite{Vodolazov_2005} and in principle it may appear if there
are defects on opposite edges of the strip with different magnetic
field controlled 'strength' leading to different suppression of
the edge barrier. The edge defects give nonreciprocal contribution
to total $L_k$ but on scale about of defect size ($\sim$ 100 nm)
along the strip which is several orders of magnitude smaller than
its length (3 mm). As a result it is difficult to observe
nonreciprocal $L(I)$ in out-of-plane field.

From comparison of Fig. 5 and 6 we conclude that if even there is
finite $B_{out}$ in the experiment with in-plane field it cannot
explain large difference between $I_c^+$ and $I_c^-$ at $B_{in}
\gtrsim 150 mT$ because $B_{out}$ suppresses them on equal foot.
Besides at $B_{out}=5 mT$ ($I_c^{\pm}(5 mT) \sim I_c^{\pm}(0)/2$)
kinetic inductance practically does not depend on current, while
in in-plane field $L$ varies with current even at $B_{in}=250 mT$
($I_c^-(250 mT) \sim I_c^{\pm}(0)/3$). Therefore we exclude
influence of $B_{out}$.

Sample-dependent $I_c^{\pm} < I_{dep}^{\pm}$ allows us to suppose
that out-of-plane vortices enter the SN strip via local
sample-dependent edge defects at $I>I_c^{\pm}$ and it launches the
resistive state - on experimental IV curves there are vortex flow
branches above $I_c^{\pm}$. At small fields, when the proximity
induced superconductivity in Cu layer is slightly suppressed by
$B_{in}$ there is proportionality $I_c\sim I_{dep}$ (coefficient
proportionality is sample-dependent). Because defect may have
variation of its properties across the thickness of MoN/Cu strip
at may lead to change the sign of diode effect, similar to one
observed in out-of-plane field. Obviously our 1D model cannot
catch this effect.

Large in-plane field stronger suppresses proximity induced
superconductivity in Cu layer. Positive current suppresses it even
more and we effectively have a single S layer with weak field
dependence of $I_c^+$ at $B_{in} \lesssim \Phi_0/d_S^2\sim 1300
mT$ (at this or larger field we expect entry of in-plane vortices
in S layer according to \cite{Shmidt_1970}) which we observe in
the experiment. Negative current partially recovers
superconductivity in N layer (it is seen from decrease of
experimental $L(I)$ at large $B_{in}$) and it may lead to
appearance of in-plane vortrices in N layer near SN interface at
field $B_{in} \sim \Phi_0/(d_S+d_N)^2 \sim 320 mT$ which is near
our experimental value. In-plane vortices should favor entry of
out-of-plane vortices and it may suppress the critical current and
change the relation between $I_c$ and $I_{dep}$. This is possible
but speculative scenario. Theoretical description of out-of-plane
vortex entry to SN strip with thickness dependent 'density' $n$,
finite momentum $q_0$ and possible existence of row of in-plane
vortices located on or close to SN interface is rather complicated
3D problem which needs separate study.

Single and multiple sign change of the diode effect with increase
of in- or out-of-plane magnetic field were observed earlier in
Refs.
\cite{Pal_Nat_Phys_2022,Sundaresh_2023,Kawarazaki_2023,Margineda_2023}
and predicted theoretically in \cite{Daido_PRL_2022,Ilic_2022} for
noncentrosymmetric superconductor. From that theories it follows
that together with sign change of SDE there is strong change of
$I(q)$ (see Fig. 5 in \cite{Daido_PRL_2022} and Fig. 2 in
\cite{Ilic_2022}) and, hence, $L_k(I)$. In our system we do not
observe drastic variation of $L_k(I)$ when sign change of diode
effect occurs and $L_k(I)$ evolves with increase of $B_{in}$ as
our theory predicts. In Refs.
\cite{Pal_Nat_Phys_2022,Sundaresh_2023,Kawarazaki_2023,Margineda_2023}
$L_k(I)$ has not been measured and it is difficult to interpret
the origin of the effect.

\subsection{Diode effect and nonreciprocal resistivity near T$_c$}

By approaching to the critical temperature $I_c^{\pm}$ decreases
but still there is diode effect and $V(I) \neq -V(-I)$ - see Fig.
7. It is known that thermal fluctuations allow vortices (we keep
in mind out-of-plane vortices) to enter the superconducting strip
at the current less than $I_c$. The probability for vortex to
overcome the edge barrier is proportional to Arrhenius factor
$exp(-dF(I)/k_BT)$ where $dF(I)$ is a current dependent height of
the edge barrier which goes to zero at $I=I_c$ and it is
proportional to the vortex energy at zero current $dF \sim
F_0=\Phi_0^2d/16\pi^2\lambda^2$
\cite{Bartolf_2010,Bulaevskii_2011,Vodolazov_2012}. At $T\sim T_c$
both $I_c$ and $F_0$ vanish which increases the impact of
fluctuations. As a result near $T_c$ the resistance $R$ is finite
even at $I\to 0$ and it could be nonreciprocal in case of
different $I_c^{\pm}$  because of $dF(I^+)\neq dF(I^-)$.

\begin{figure}[hbtp]
\includegraphics[width=0.48\textwidth]{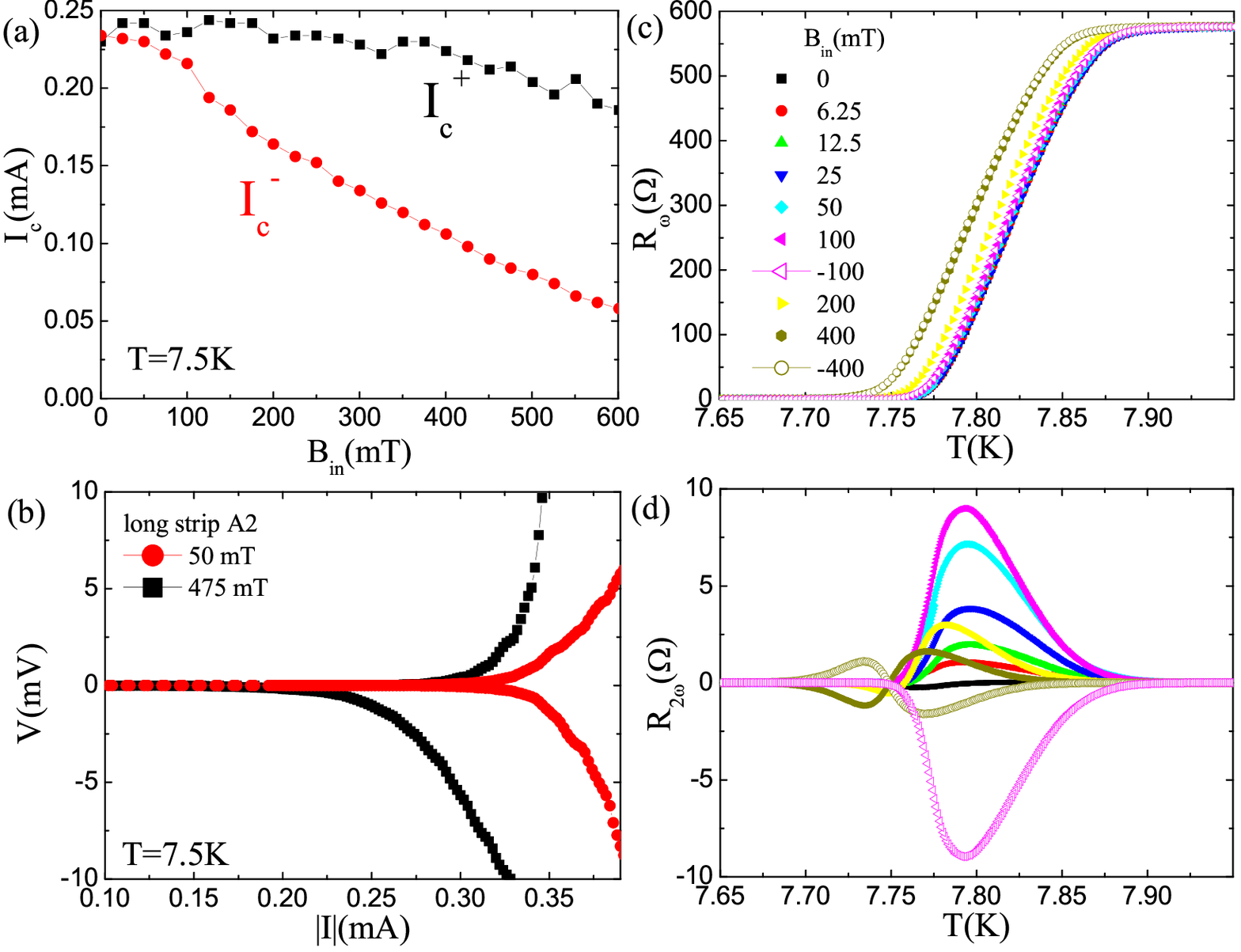}
\caption{Panel (a) - field dependent critical currents $I_c^{\pm}$
of sample A2 at T=7.5K. Panel (b) - current voltage
characteristics of strip A2 at different $B_{in}$. Panels (c,d) -
temperature dependence of first and second harmonics of IV
characteristics at different in-plane magnetic fields.}
\end{figure}

Additional mechanism which may contribute to nonreciprocal $R$ is
the current dependent viscosity of vortex motion. Indeed, the
transport current with one direction stronger suppresses proximity
induced superconductivity in N layer than the current with
opposite direction which should affect the vortex velocity $v$ and
voltage $V\sim v$. Our measurements of the first and second
harmonics of IV characteristics confirm existence of nonreciprocal
resistance at finite $B_{in}$ near $T_c$ (see panels c,d in Fig.
7).

Earlier finite $R_{2\omega}$ near $T_c$ has been found for various
hybrid superconducting structures placed in in-plane magnetic
field where its origin has been related with spin-orbit coupling
\cite{Wakatsuki_SciAdv_2017,Qin_Nat_Comm_2017,Yasuda_Nat_Comm_2019,Ando_Nat_2020}.
Our experiment demonstrates existence of qualitatively similar
result in MoN/Cu hybrid where we have orbital mechanism of finite
momentum state, resulting in diode effect and nonreciprocal
resistance near critical temperature. We believe that the same
orbital mechanism has to exist in Ref.
\cite{Ando_Nat_2020,Wakatsuki_SciAdv_2017,Qin_Nat_Comm_2017,Yasuda_Nat_Comm_2019}
and compete with spin-orbit interaction.

\section{Discussion}

Proposed orbital mechanism of finite momentum state in hybrid
superconductors is finite size (thickness) effect. We
theoretically find that at relatively large $B_{in}$ $q_{NCS} \sim
-B_{in}(d_S+d_N)2\pi/\Phi_0$ for different thicknesses of S and N
layers (for parameters of SN bilayer from Fig. 3 $q_{NCS} \simeq -
B_{in}(d_S+d_N)2\pi/3\Phi_0$). It means that when the thickness of
SN bilayer goes to zero $q_{NCS} \to 0$ at and orbital mechanism
disappears (the same is valid when $d_N \to 0$). It gives the way
to check its influence in the experiment - one may study thickness
dependence of $L_k(I,B_{in})$ because for smaller thickness the
larger $B_{in}$ is needed to have the same $q_{NCS}$. Indeed, we
find that for two times thinner sample MoN(20nm)/Cu(20nm) the
change of $L(I)$ is less sensitive to $B_{in}$ (compare Fig. 10(a)
with Fig. 5(a,b)) than for thick strip which qualitatively
coincide with our calculations (compare Fig. 3(b,d) with Fig.
10(c,d)).

Our experimental results indicate that one should be careful with
interpretation of experiments on diode effect. We observe sign
change of the diode effect with increase of magnetic field in two
(one long and one short) thick samples and its absence in another
two long and short thick samples at $B_{in} \lesssim 150 mT$ while
$L(I, B_{in})$ and, hence, $I(q)$ was almost the same for all long
samples. It means that additional factor, having not intrinsic to
given material character, may play a role. In our case we believe
that sample-dependent edge defects are responsible for observed
effect. In this respect measurements of $L_k(I,B_{in})$ give more
reliable information about intrinsic origin of finite momentum
state than measurements of $I_c^{\pm}(B_{in})$. Indeed, if there
is strong variation of superconducting properties along the sample
but on small distance it weakly affects the total kinetic
inductance but it may strongly affect critical current because it
is determined by the 'weakest' place.

At $B_{in} \gtrsim 150 mT$ for all thick samples we find sign
change of the diode effect and its considerably larger value in
comparison with theoretical expectations. At the same time we do
not observe any qualitative changes of $L(I)$. We believe that it
could be connected with appearance of in-plane vortices in SN
bilayer. If it is true it is also finite thickness effect and with
decreasing the thickness of SN bilayer in-plane vortices (and sign
change of diode effect) should appear at larger field. Somewhere
similar effect appears in thin strip where we do not have sign
change of diode effect at 2.7 K for all studied three samples up
to $B_{in}=500 mT$ and at $4.6 K$ it exists only for one sample.

In Ref. \cite{Ando_Nat_2020} the diode effect was observed in
multilayered Nb/V/Ta strip with thickness $d=120 nm$ and width
$w=50 \mu m$. At low temperature SDE vanished which is in contrast
to results of Refs.
\cite{Turini_N_Lett_2022,Baumgartner_Nat_Nan_2022,Pal_Nat_Phys_2022,Sundaresh_2023}
and our work where diode effect becomes more pronounced at low T.
The width of Nb/V/Ta strip greatly exceeds effective Pearl
magnetic field penetration depth $\lambda^2/d$ except at $T\sim
T_c$ (we assume that $\lambda(0) =120 nm$ as in dirty Nb because
zero temperature coherence length $\xi(0)=13 nm$ of multilayer is
close to $\xi$ of dirty Nb). It means that current distribution is
nonuniform over the width of Nb/V/Ta strip which automatically
means that critical current has to be much smaller than the
depairing current and resistive state is connected with entry and
motion of out-of-plane vortices. The nonuniform current
distribution may lead to vortex pinning at low temperature which
destroys experimentally observed diode effect.

In Al/InGaAs/InAs/InGaAs heterostructures SDE was found and
explained by the interplay between diamagnetic and external
currents \cite{Sundaresh_2023}, which qualitatively coincides with
mechanism of the diode effect in SN bilayer proposed in Ref.
\cite{Vodolazov_2018}, here and in Ref. \cite{Vodolazov_2005} for
the superconducting strip with nonequivalent edges being in
out-of-plane magnetic field. Assumption of authors of
\cite{Sundaresh_2023} that different layers in heterostructure
have different currents may be reformulated in terms of different
densities of superconducting electrons and finite $\nabla n$ which
has to lead to finite $q_{NCS}$ when there is in-plane magnetic
field. In terms of Ref. \cite{Sundaresh_2023} we have strong
coupling regime between S and N layers and in our case if in-plane
vortices appear at large $B_{in}$ they should be more like
Abrikosov vortices than Josephson ones.

In Refs.
\cite{Turini_N_Lett_2022,Baumgartner_Nat_Nan_2022,Pal_Nat_Phys_2022}
SDE was observed in SNS Josephson junction (JJ) and, hence, our
results cannot be applied directly to that systems. But all
studied junctions have hybrid SN banks where in-plane magnetic
field should produce FMS. There is a question, may finite
$q_{NCS}$ in the SN banks affects the transport properties of JJ
even if in N weak link there is no $\nabla n$ or it is small? To
answer this question one has to calculate transport properties of
SN-N-SN junction taking into account finite thicknesses of SN
banks and normal weak link which is difficult 2D problem.
Therefore at the moment we cannot claim that the orbital mechanism
is involved in diode effect observed in that works, although
results from Refs.
\cite{Turini_N_Lett_2022,Baumgartner_Nat_Nan_2022,Pal_Nat_Phys_2022}
look qualitatively similar to our results.

\section{Conclusion}

We demonstrate appearance of finite momentum state in SN bilayer
placed in in-plane magnetic field. Experimentally, presence of FMS
is proven via observation of nonreciprocal inductance $L(I)\neq
L(-I)$ in several MoN/Cu strips being in in-plane magnetic field.
It has an orbital nature as it follows from the experiment with
samples having different thickness and expected zero or negligible
contribution of spin-orbit coupling in our system.

We also experimentally observe superconducting diode effect but in
contrast to results with $L(I,B_{in})$ it has rather cumbersome
sample-dependent behavior. We speculate that it could be connected
with presence of the sample-dependent edge defects and appearance
of in-plane vorticies in large enough $B_{in}$.

Taking into account our results and discussion in Introduction we
may claim that finite momentum state is not elusive or rear
phenomena in superconducting structures. Any nonuniformities
(material or geometric) lead to $\nabla n \neq 0$ and in presence
of magnetic field the particular direction appears along which
there is a difference in critical currents (SDE) and nonreciprocal
$L_k$. If the diode effect originates from local edge defects (as
in case of superconducting strip placed in out-of-plane field) it
may not lead to noticeable nonreciprocal $L_k$ of a whole sample
due to local nature of FMS in this case.

\begin{acknowledgments}

The work is supported by the Russian Science Foundation (project
No. 23-22-00203).
\end{acknowledgments}

\appendix

\section{Model}

To calculate transport properties (critical current, kinetic
inductance) of SN bilayer we use the one-dimensional Usadel
equation for normal $g$ and anomalous $f$ quasi-classical Green
functions \cite{Usadel}. With standard angle parametrization
$g=cos\Theta$ and $f=sin\Theta\exp(i\varphi)$ the Usadel equations
in different layers can be written as
\begin{equation}
\label{usadel-s} \hbar D_S\frac{\partial^2\Theta_S}{\partial
z^2}-\left(\hbar\omega_k+D_S\hbar q^2\cos
\Theta_S\right)\sin\Theta_S+2\Delta\cos\Theta_S=0,
\end{equation}

 \begin{equation}
  \label{usadel-n}
  \hbar D_N\frac{\partial^2\Theta_N}{\partial
z^2}-\left(\hbar\omega_k+D_N\hbar q^2\cos
\Theta_N\right)\sin\Theta_N=0,
\end{equation}
where subscripts S and N refer to superconducting and normal
layers, respectively. Here $D$ is the diffusion coefficient for
corresponding layer, $\hbar \omega_k = \pi k_BT(2k+1)$ are the
Matsubara frequencies (k is an integer number), $\hbar
q=\hbar(\nabla\varphi + 2\pi\,{\bf A}/\Phi_0)=\hbar(q_0 +
2\pi\,{\bf A}/\Phi_0)$ is the momentum of Cooper pairs, $\varphi$
is the phase of the order parameter, ${\bf A}$ is the vector
potential, $\Phi_0=\pi\hbar c/|e|$ is the magnetic flux quantum.
$\Delta$ is the superconducting order parameter, which satisfies
to the self-consistency equation
\begin{equation}
\label{self-cons} \Delta \ln\left(\frac{T}{T_{c0}}\right)=2\pi k_B
T\sum_{\omega_k
>0} \left(\sin\Theta_S - \frac{\Delta}{\hbar\omega_k}\right),
\end{equation}
where $T_{c0}$ is the critical temperature of single S layer in
the absence of magnetic field. These equations are supplemented by
the Kupriyanov-Lukichev boundary conditions on SN interface
\cite{JETP-1988}
  \begin{equation}
    D_S\frac{d\Theta_S}{dz}=D_N\frac{d\Theta_N}{dz}
   \end{equation}
For simplicity we consider case with zero barrier between layers
and continuous $\Theta$ on SN interface. For interfaces with
vacuum we use the boundary condition $d\Theta/dz=0$.

We assume that the thickness $d_S+d_N$ of SN strip is much smaller
than the London penetration depth $\lambda$ while the width $w$ is
smaller than the Pearl penetration depth
$\Lambda=\lambda^2/(d_S+d_N)$ which allows us to neglect the
effect of superconducting screening on the vector potential and
magnetic field. We choose vector potential ${\bf
A}=(-B_{in}z,0,0)$ with thickness averaged $\int {\bf A} dz=0$
(here $-(d_S+d_N)/2<z<(d_S+d_N)/2$).

To calculate the current density $j$, 'density' of superconducting
electrons $n$ and current $I=w\int jdz$ we use the standard
expression
\begin{eqnarray}
j(z)=-\frac{2\pi k_BT}{|e|\rho}q(z)\sum_{\omega_k >
0}\sin^2\Theta=-|e|n(z)q(z)\frac{\hbar}{m}
\\
n(z)=\frac{m}{\hbar}\frac{2\pi k_BT}{e^2\rho}\sum_{\omega_k >
0}\sin^2\Theta=\frac{mc^2}{8\pi |e|\lambda^2(z)}
\end{eqnarray}
where $\rho=2|e|D_{S,N}N(0)$ is the residual resistivity of the
corresponding layer, $N(0)$ is density of states of electrons per
one spin at the Fermi level in the normal state (we assume
identical $N(0)$ in S and N layers) and $m$ is a 'mass' of
superconducting electrons.
\begin{figure}[hbtp]
\includegraphics[width=0.48\textwidth]{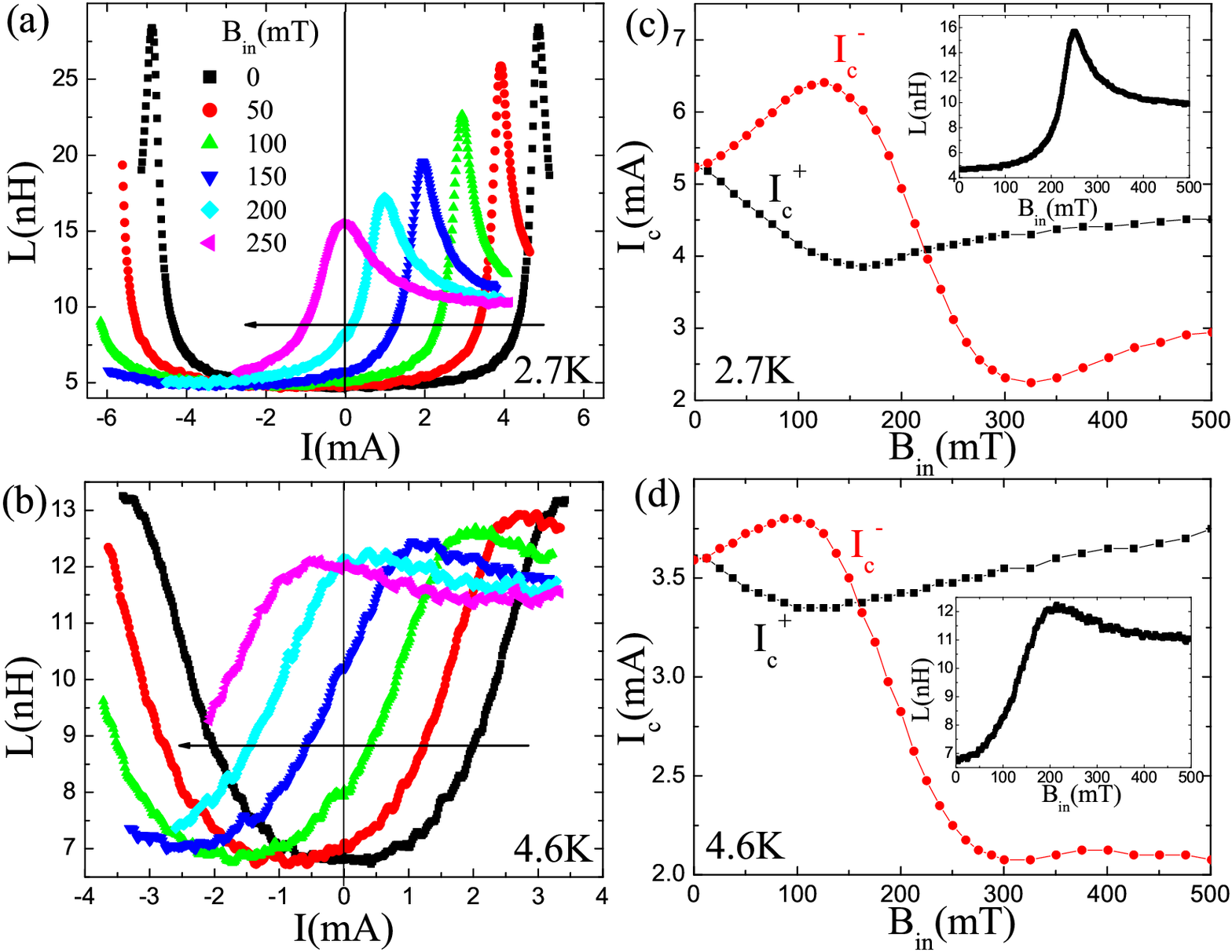}
\caption{Panels (a,b) - evolution of current dependent inductance
of MoN/Cu strip (sample A4) with increase of in-plane magnetic
field at T=2.7K and T=4.6K, respectively. In panels (c,d) we
present field dependent critical current and inductance.}
\end{figure}

Equations (A1-A3) are solved numerically by using iteration
procedure. For initial distribution $\Delta(z)=const$ and chosen
$q_0$, $B_{in}$ we solve Eqs. (A1,A2) (in numerical procedure we
use Newton method combined with tridiagonal matrix algorithm).
Found solution $\Theta(z)$ is inserted to Eq. (A3) to find
$\Delta(z)$ and than iterations repeat until the relative change
in $\Delta(z)$ between two iterations does not exceed $10^{-8}$.
Length is normalized in units of $\xi_c=\sqrt{\hbar
D_S/k_BT_{c0}}$, energy is in units of $k_BT_{c0}$, current is in
units of depairing current $I_{dep,S}$ of single S layer with the
thickness $d_S$, the magnetic field is in units of
$B_0=\Phi_0/2\pi\xi_c^2$ ($B_0$ by factor 1.76 is smaller than the
out-of plane second critical field $B_{c2}(T=0)$ of single S
layer). Typical step grid in S and N layers is $\delta z=0.1
\xi_c$. In calculations we used the following parameters:
$d_S=d_N=4\xi_c$, $D_N/D_S=100$ for thick strip and
$d_S=d_N=2\xi_c$, $D_N/D_S=80$, which are not far from
experimental values.
\begin{figure}[hbtp]
\includegraphics[width=0.45\textwidth]{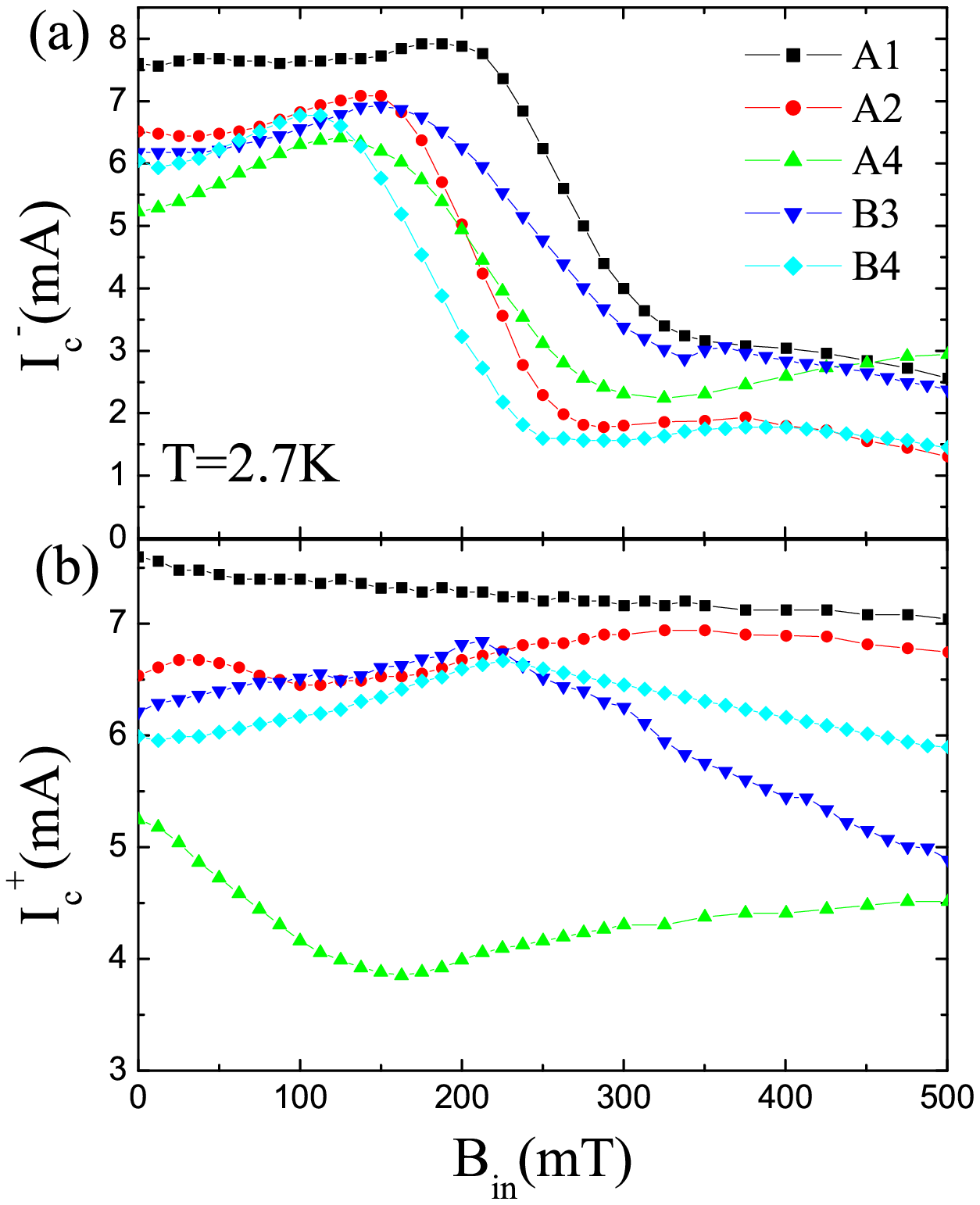}
\caption{Panels (a) and (b)- field dependent critical currents
$I_c^{\pm}$ for different MoN/Cu strips (A1, A2, A4 - long strips,
B3, B4 - short strips) at $T=2.7K$.}
\end{figure}

With calculated $I(q_0)$ we find kinetic inductance per unit of
length of the strip
\begin{equation}
L_k=-\hbar c^2(dI/dq_0)^{-1}/2|e|
\end{equation}
and $I_{dep}^{\pm}$ as {\it absolute} value of maximal positive
and negative superconducting currents where $dI/dq_0=0$.

\section{Results for different thick samples}

In Fig. 8 we show $L(I,B_{in})$ and $I_c^{\pm}(B_{in})$ for strip
A4. While dependence $L(I,B_{in})$ is rather close to one for
strip A2 (the same is valid for sample A1 - results are not shown
here) field dependencies of $I_c^{\pm}$ are quantitatively
different at $B_{in} \lesssim 150 mT$ and $T=2.7K$. We explain it
by specific to each sample edge defects.

In Fig. 9 we show sample-dependent $I_c^{\pm}(B_{in})$ for all
studied three long and two short strips at $T=2.7K$. Despite of
quantitative differences one may notice some general properties
for all strips. At weak fields ($B_{in}\lesssim 150 mT$) $I_c^-$
slightly increases, which is in accordance with our theory. But
than there is sharp decrease of $I_c^-$ and it becomes much
smaller than $I_c^+$, which is opposite to our model. On contrary,
critical current $I_c^+$ varies much weaker while its field
dependence changes from strip to strip.

\section{Results for thinner samples}

In Fig. 10(a,b) we present results of transport measurements for
two times thinner MoN/Cu long strips at $T=4.6 K$ (samples A1-A3,
the width and length are the same as for thick long strips,
$d_{MoN}=d_{Cu}=20nm$, $T_c=7.37K$, 20 nm thick Cu has $\rho=2.9
\mu Ohm \cdot cm$ at T=10 K, resistivity of MoN layer is the same
as for 40 nm thick layer). In Fig. 10 (c,d) we show theoretical
results ($d_S=d_N=2\xi_c$, $D_N/D_S=80$, $T=0.6 T_{c0}$).
Inductance is nonreciprocal in presence of in-plane field which is
proof of appearance of finite momentum state but we need much
larger field to have qualitatively the same change of $L$ as for
thick strip. From the experiment we conclude that our maximal
accessible field ($B_{in}=500 mT$) does not destroy proximity
induced superconductivity in Cu layer (at zero current) because we
do not see a peak in dependence $L(B_{in})$ (see inset in Fig. 10
(b)). Qualitatively the same results are obtained at $T=2.7 K$
(not shown here). These findings prove that we have in the
experiment orbital mechanism of finite momentum state.

Sign change of the diode effect is observed only in sample A1 (see
Fig. 10(b)) at 4.6K and for all studied samples it does not exist
at 2.7K (dependencies $I_c^{\pm}(B_{in})$ resemble ones for
samples A2-A3 at $T=4.6K$ with maximum of $I_c^-(B_{in})$ at $\sim
250 mT $). In this respect results for thinner sample better
follow our theoretical calculations with no sign change of the
diode effect (see inset in Fig. 10(c)). However we cannot rule out
its appearance at large fields where $L$ should reach the peak in
dependence $L(B_{in})$.

\begin{figure}[hbtp]
\includegraphics[width=0.48\textwidth]{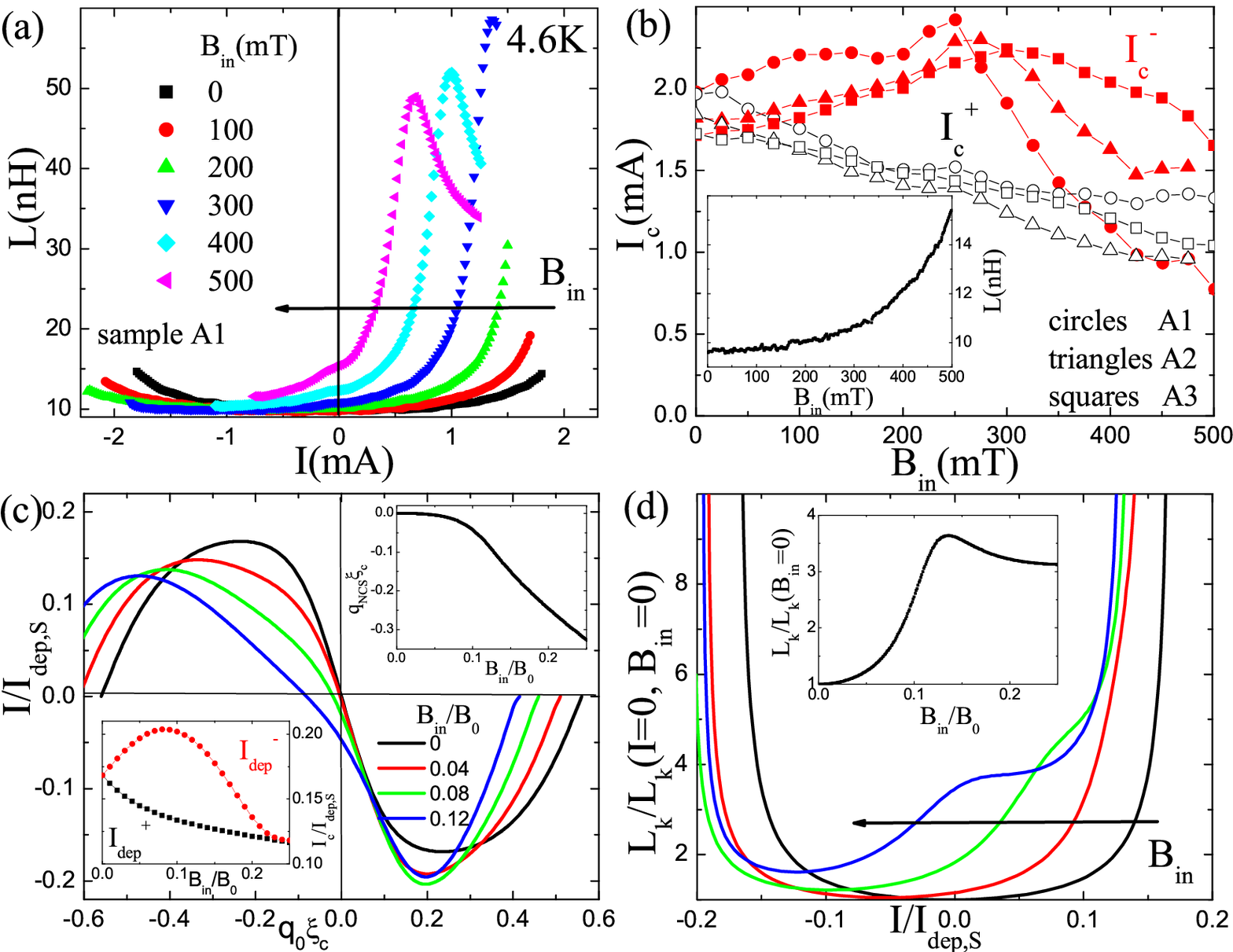}
\caption{Panel (a) - evolution of current dependent inductance of
thin MoN(20nm)/Cu(20nm) strip (sample A1) with increase of
in-plane magnetic field at T=4.6K. Panel (b) - field dependent
positive and negative critical currents of strips A1-A3 at T=4.6
K. In panels (c,d) we present results of theoretical calculations
for thin SN strip.}
\end{figure}


\begin{references}

\bibitem{Fulde_1964} P. Fulde and R. A. Ferrell, Superconductivity in a strong
spin-exchange field, Phys. Rev. {\bf 135}, A550 (1964).

\bibitem{Mironov_2012} S. Mironov, A. Mel'nikov, and A. Buzdin, Vanishing Meissner
effect as a Hallmark of in-Plane Fulde-Ferrell-Larkin-Ovchinnikov
Instability in Superconductor-Ferromagnet Layered Systems, Phys.
Rev. Lett. {\bf 109}, 237002 (2012).

\bibitem{JETPL-2014} A. M. Bobkov and I. V. Bobkova, Enhancing of the Critical
Temperature of an In-Plane FFLO State in Heterostructures by the
Orbital Effect of the Magnetic Field, JETP Letters, {\bf 99}, 333
(2014).

\bibitem{Mironov_2018} S. V. Mironov, D. Vodolazov, Yu. Yerin, A. V. Samokhvalov,
A. S. Melnikov, and A. Buzdin, Temperature Controlled
Fulde-Ferrell-Larkin-Ovchinnikov Instability in
Superconductor-Ferromagnet Hybrids, Phys. Rev. Lett. {\bf 121},
077002 (2018).

\bibitem{Bobkova_2013} I. V. Bobkova and A. M. Bobkov, In-plane
Fulde-Ferrel-Larkin-Ovchinnikov instability in a
superconductor/normal metal bilayer system under nonequilibrium
quasiparticle distribution, Phys. Rev. B {\bf 88}, 174502 (2013).

\bibitem{Ouassou_2020} J. A. Ouassou, W. Belzig, and J. Linder,
Prediction of a Paramagnetic Meissner Effect in Voltage-Biased
Superconductor-Normal-Metal Bilayers, Phys. Rev. Lett., {\bf 124},
047001 (2020).

\bibitem{Levichev_2021} M. Yu. Levichev, I. Yu. Pashenkin, N. S. Gusev, and D. Yu. Vodolazov,
Voltage controllable superconducting state in the multiterminal
superconductor-normal-metal bridge, Phys. Rev. B {\bf 103}, 174507
(2021).

\bibitem{Plastovets_2019} V. D. Plastovets  and D. Y. Vodolazov,
Dynamics of Domain Walls in a Fulde-Ferrell Superconductor, JETP
Lett. {\bf 109}, 729 (2019).

\bibitem{Samokhin_2017} K. V. Samokhin, B. P. Truong, Current-carrying
states in Fulde-Ferrell-Larkin-Ovchinnikov superconductors, Phys.
Rev. B {\bf 96}, 214501 (2017).

\bibitem{Marychev_2021} P.M. Marychev and D.Y. Vodolazov, Extraordinary kinetic inductance
of superconductor/ferromagnet/normal metal thin strip in an
Fulde-Ferrell state, Journal of Physics, Condensed Matter {\bf
33}, 385301 (2021).

\bibitem{Edelshtein_JETP_1989} V.M. Edelshtein, Characteristics of the Cooper pairing in two-dimensional noncentrosymmetric electron
systems, Sov. Phys.-JETP {\bf 68}, 1244 (1989).

\bibitem{Agterberg_2003} D.F. Agterberg, Novel magnetic field effects in unconventional
superconductors, Physica C {\bf 387}, 13 (2003).

\bibitem{book_2012} Non-Centrosymmetric Superconductors, edited by E. Bauer and M.
Sigrist (Springer, Berlin, 2012).

\bibitem{Yuan_PNAS_2022} N. F. Q. Yuan and L. Fu, Supercurrent diode effect and
finite-momentum superconductors, Proceedings of the National
Academy of Sciences {\bf 119} (2022), 10.1073/pnas.2119548119.

\bibitem{Daido_PRL_2022} A. Daido, Y. Ikeda, Y. Yanase, Intrinsic superconducting diode
effect, Phys. Rev. Lett. {\bf 128}, 037001 (2022).

\bibitem{He_NJP_2022} J. J. He, Y. Tanaka, N. Nagaosa, A phenomenological theory of superconductor
diodes, New J. Phys. {\bf 24}, 053014 (2022).

\bibitem{Baumgartner_Nat_Nan_2022} C. Baumgartner, L. Fuchs, A. Costa, S. Reinhardt, S. Gronin, G.
C. Gardner, T. Lindemann, M. J. Manfra, P. E. F. Junior, D.
Kochan, J. Fabian, N. Paradiso, C. Strunk, Supercurrent
rectification and magnetochiral effects in symmetric Josephson
junctions, Nat. Nanotech. {\bf 17}, 39 (2022).

\bibitem{Vodolazov_2018} D. Yu. Vodolazov, A. Yu. Aladyshkin , E. E. Pestov, S. N.
Vdovichev, S. S. Ustavshikov, M. Yu. Levichev, A. V. Putilov, P.
A. Yunin, A. I. El'kina, N. N. Bukharov and A. M. Klushin,
Peculiar superconducting properties of a thin film
superconductor-normal metal bilayer with large ratio of
resistivities, Supercond. Sci. Technol. {\bf 31}, 115004 (2018).

\bibitem{Ando_Nat_2020} F. Ando, Y. Miyasaka, T. Li, J. Ishizuka, T.
Arakawa, Y. Shiota, T. Moriyama, Y. Yanase, T.Ono, Observation of
superconducting diode effect, Nature {\bf 584}, 373 (2020).

\bibitem{Turini_N_Lett_2022} B. Turini, S. Salimian, M. Carrega, A. Iorio, E. Strambini, F.
Giazotto, V. Zannier, L. Sorba, and S. Heun, Josephson Diode
Effect in High-Mobility InSb Nanoflags, Nano Lett. {\bf 22}, 8502
(2022).

\bibitem{Pal_Nat_Phys_2022} B. Pal, A. Chakraborty, P. K. Sivakumar, M. Davydova, A. K. Gopi, A. K.
Pandeya, J. A. Krieger, Y. Zhang, M. Date, S. Ju, N. Yuan, N. B.
M. Schroter, L. Fu and S. S. P. Parkin, Josephson diode effect
from Cooper pair momentum in a topological semimetal, Nature Phys.
{\bf 18}, 1228 (2022).

\bibitem{Sundaresh_2023} A. Sundaresh, J. I. Vayrynen, Y. Lyanda-Geller and L. P.
Rokhinson, Diamagnetic mechanism of critical current
non-reciprocity in multilayered superconductors, Nat. Commun. {\bf
14}, 1628 (2023).

\bibitem{Vodolazov_2005} D. Y. Vodolazov and F. M. Peeters, Superconducting rectifier
based on the asymmetric surface barrier effect, Phys. Rev. B {\bf
72}, 172508 (2005).

\bibitem{Greenhouse} H. M. Greenhouse, Design of Planar Rectangular Microelectronic
Inductors,  IEEE Trans. Parts, Hybrids, Packag. {\bf 10}, 101
(1974).

\bibitem{JETP_Ustavschikov} S. Ustavschikov, M. Y. Levichev, I. Y. Pashenkin, N. Gusev, S.
Gusev, and D. Y. Vodolazov, Diode Effect in a Superconducting
Hybrid Cu/MoN Strip with a Lateral Cut, J. Exp. Theor. Phys. {\bf
135}, 226 (2022).

\bibitem{Cerbu_NJP_2013} D. Cerbu, V. N. Gladilin, J. Cuppens, J. Fritzsche,
J. Tempere, J. T. Devreese, V. V. Moshchalkov, A. V. Silhanek, and
J. Van de Vondel, Vortex ratchet induced by controlled edge
roughness, New J. Phys. {\bf 15}, 063022 (2013).

\bibitem{PRB14Ilin} K. Ilin, D. Henrich, Y. Luck, Y. Liang, and M. Siegel,
D. Yu. Vodolazov, Critical current of Nb, NbN, and TaN thin-film
bridges with and without geometrical nonuniformities in a magnetic
field, Phys. Rev. B {\bf 89}, 184511 (2014).

\bibitem{Suri_APL_2022} D. Suri, A. Kamra, T. N. G. Meier,M. Kronseder, W.
Belzig, Ch. H. Back, and Ch. Strunk, Non-reciprocity of
Vortex-limited Critical Current in Conventional Superconducting
Micro-bridges, Appl. Phys. Lett. {\bf 121}, 102601 (2022).

\bibitem{Hou_2022} Y. Hou, F. Nichele, H. Chi, A. Lodesani, Y. Wu, M.F. Ritter, D.
Z. Haxell, M. Davydova, S. Ilic, O. Glezakou-Elbert, A.
Varambally, F. S. Bergeret, A. Kamra, L. Fu, P. A. Lee, J. S.
Moodera, Ubiquitous Superconducting Diode Effect in Superconductor
Thin Films, Phys. Rev. Lett. {\bf 131}, 027001 (2023).

\bibitem{Satchell} N. Satchell, P.M. Shepley, M.C. Rosamond, and G.
Burnell, Supercurrent diode effect in thin film Nb tracks, J. of
Appl. Phys. {\bf 133}, 203901 (2023).

\bibitem{Bauriedl_Nat_Comm_2022} L. Bauriedl, Ch. Bauml, L. Fuchs, Ch. Baumgartner, N.
Paulik, J. M. Bauer, K.-Q. Lin, J. M. Lupton, T. Taniguchi, K.
Watanabe, Ch. Strunk, and  N. Paradiso, Nat. Commun. {\bf 13},
4266 (2022).

\bibitem{Plourde_PRB_2001} B. L. T. Plourde, D. J. Van Harlingen, D. Yu. Vodolazov, R.
Besseling, M. B. S. Hesselberth, and P. H. Kes, Influence of edge
barriers on vortex dynamics in thin weak-pinning superconducting
strips, Phys. Rev. B {\bf 64}, 014503 (2001).

\bibitem{Shmidt_1970} V.V. Shmidt, The critical current in superconducting films, Sov. Phys.-JETP {\bf 30}, 1137 (1970).

\bibitem{Kawarazaki_2023} R. Kawarazaki, R. Iijima, H. Narita, R. Hisatomi, Y. Shiota, T.
Moriyama, and T. Ono, Rectification effect of non-centrosymmetric
Nb/V/T superconductor, Journal of the Magnetics Society of Japan,
2309R001 (2023).

\bibitem{Margineda_2023} D. Margineda, A. Crippa, E. Strambini, Y. Fukaya, M. T.
Mercaldo, M. Cuoco, and F. Giazotto, Sign reversal diode effect in
superconducting Dayem nanobridges, arXiv:2306.00193.

\bibitem{Ilic_2022} S. Ilic and F. S. Bergeret, Theory of the supercurrent diode effect in Rashba superconductors
with arbitrary disorder, Phys. Rev. Lett. {\bf 128}, 177001
(2022).

\bibitem{Bartolf_2010} H. Bartolf, A. Engel, A. Schilling, K. Ilin,M. Siegel, H.-W.
Hubers, and A. Semenov, Current-assisted thermally activated flux
liberation in ultrathin nanopatterned NbN superconducting meander
structures, Phys. Rev. B  {\bf 81}, 024502 (2010).

\bibitem{Bulaevskii_2011} L. N. Bulaevskii, M. J. Graf, C. D. Batista, and V. G. Kogan,
Vortex-induced dissipation in narrow current-biased thin-film
superconducting strips, Phys. Rev. B {\bf 83}, 144526 (2011).

\bibitem{Vodolazov_2012} D. Y. Vodolazov, Saddle point states in two-dimensional
superconducting films biased near the depairing current, Phys.
Rev. B {\bf 85}, 174507 (2012).

\bibitem{Wakatsuki_SciAdv_2017} R. Wakatsuki, Yu Saito, S. Hoshino,  Y. M. Itahashi,  T. Ideue, M.
Ezawa, Y. Iwasa, N. Nagaosa, Nonreciprocal charge transport in
noncentrosymmetric superconductors, Sci. Adv. {\bf 3}, e1602390
(2017).

\bibitem{Qin_Nat_Comm_2017} F. Qin, W. Shi, T. Ideue, M. Yoshida, A. Zak, R. Tenne, T. Kikitsu, D. Inoue, D.
Hashizume, Y. Iwasa, Superconductivity in a chiral nanotube, Nat.
Commun. {\bf 8}, 14465 (2017).

\bibitem{Yasuda_Nat_Comm_2019} K. Yasuda, H. Yasuda, T. Liang, R. Yoshimi, A. Tsukazaki, K. S.
Takahashi, N. Nagaosa, M. Kawasaki, and Y. Tokura, Nonreciprocal
charge transport at topological insulator/superconductor
interface, Nat. Commun. {\bf 10}, 2734 (2019).

\bibitem{Usadel} K. D. Usadel, Generalized diffusion equation for superconducting alloys, Phys. Rev. Lett. {\bf 25}, 507 (1970).

\bibitem{JETP-1988} M. Yu. Kuprianov and V. F. Lukichev, Influence of boundary
transparency on the critical current of "dirty" SS'S structures,
Sov. Phys. JETP {\bf 67}, 1163 (1988).

\end{references}
\end{document}